\documentclass[conference]{IEEEtran}
\IEEEoverridecommandlockouts
\usepackage{cite}
\usepackage{amsmath,amssymb,amsfonts}
\usepackage{algorithmic}
\usepackage{graphicx}
\usepackage{textcomp}
\usepackage{xcolor}
\usepackage{comment}
\usepackage{hyperref}
\usepackage{amssymb}
\usepackage{bm}
\usepackage{afterpage}
\usepackage{lmodern}
\def\BibTeX{{\rm B\kern-.05em{\sc i\kern-.025em b}\kern-.08em
    T\kern-.1667em\lower.7ex\hbox{E}\kern-.125emX}}

\begin{document}

\title{
Learning Sentinel-2 reflectance dynamics for data-driven 
 assimilation and forecasting\\
\thanks{This work was supported by Agence Nationale de la Recherche under grant ANR-21-CE48-0005 LEMONADE.}
}
\author{\IEEEauthorblockN{Anthony Frion}
\IEEEauthorblockA{
\textit{Lab-STICC, IMT Atlantique} \\
Brest, France \\
anthony.frion@imt-atlantique.fr}
\and
\IEEEauthorblockN{Lucas Drumetz}
\IEEEauthorblockA{
\textit{Lab-STICC, IMT Atlantique} \\
Brest, France \\
lucas.drumetz@imt-atlantique.fr}
\and
\IEEEauthorblockN{Guillaume Tochon}
\IEEEauthorblockA{
\textit{LRE EPITA} \\
 Le Kremlin-Bicêtre, France \\
guillaume.tochon@lrde.epita.fr}
\and
\IEEEauthorblockN{Mauro Dalla Mura}
\IEEEauthorblockA{\textit{Univ. Grenoble Alpes, CNRS, Grenoble INP, GIPSA-lab} \\
\textit{Institut Universitaire de France}\\
Grenoble, France \\
mauro.dalla-mura@gipsa-lab.grenoble-inp.fr}
\and
\IEEEauthorblockN{Abdeldjalil Aissa El Bey}
\IEEEauthorblockA{\textit{Lab-STICC} \\
\textit{IMT Atlantique}\\
Brest, France \\
abdeldjalil.aissaelbey@imt-atlantique.fr}
}

\maketitle

\begin{abstract}
Over the last few years, massive amounts of satellite multispectral and hyperspectral images covering the Earth's surface have been made publicly available for scientific purpose, for example through the European Copernicus project. 
Simultaneously, the development of self-supervised learning (SSL) methods has sparked great interest in the remote sensing community, enabling to learn latent representations from
unlabeled data to help treating downstream tasks for which there is few annotated examples, such as interpolation, forecasting or unmixing.
Following this line, we train a deep learning model inspired from the Koopman operator theory to model long-term reflectance dynamics in an unsupervised way. We show that this trained model, being differentiable, can be used as a prior for data assimilation in a straightforward way. Our datasets, which are composed of Sentinel-2 multispectral image time series, are publicly released with several levels of treatment.
\end{abstract}

\begin{IEEEkeywords}
Self-supervised learning, Sentinel-2, satellite image time series, Koopman operator, Data assimilation
\end{IEEEkeywords}

\section{Introduction}




Longstanding problems in satellite image time series processing include change detection \cite{ChangeDetection}, content classification \cite{Breizhcrops}, semantic segmentation \cite{TimeSen2Crop} and spectral unmixing \cite{SpectralUnmixing}. 
In this paper, we approach these issues in a holistic way, in a self-supervised learning (SSL) context. Indeed, we design a machine learning model first trained on a pretext task without using any annotations, and \emph{in fine} use its learnt latent representation to handle downstream tasks, possibly with some labels. Our pretext task is to predict the long-term reflectance of a pixel using a given initial condition. We aim at learning discrete dynamical systems written in a generic way as 
\begin{equation}
    x_{t+1} = f(x_t;\theta)
\end{equation}
where $x$ is an observed time series and $\theta$ represents underlying parameters.  While SSL has been extensively studied for remote sensing \cite{SSL_RS_review}, to our knowledge, our work is the first to use temporal prediction as a pretext task. Our resulting model is well aware of the reflectance dynamics and can serve multiple time-related purposes, like
interpolation, denoising or forecasting. 
Its differentiability and small number of parameters makes it more versatile than many model-driven priors for downstream tasks that can be formulated as optimization problems.
In spirit, our learning approach is related to recent advances in natural language processing, e.g. \cite{GPT-3}, where a large language model 
is simply trained to predict the data and can then be asked to perform a variety of tasks.

Our contributions include: 
(1) we adapt a neural architecture that we previously introduced in \cite{ICASSP}, which learns the behavior of dynamical systems from observation data, to real-world satellite image time series
and study tools to leverage the spatial structure of these data,
(2) we show how to use such a trained model for data assimilation, in settings with sparse and irregular available data, showing promising potential to design efficient gap-filling algorithms for such remote sensing datasets,
(3) we collect, clean and interpolate two long Sentinel-2 time series, which we publicly share
(\url{https://github.com/anthony-frion/Sentinel2TS})
to make it easier for the interested community to work on similar tasks and compare their results to ours.
\section{Our methods}
\label{Method}

Our approach to learning time series dynamics is based on the Koopman operator theory \cite{Koopman}. In short, this theory states that any given dynamical system can be described by a linear operator which is applied to observation functions of the system. However, this operator, which is called the Koopman operator, is generally infinite dimensional. We refer the reader to \cite{KoopmanReview} for a recent review on this theory. Our method follows a line opened by \cite{KoopmanKIS} which aims at finding a Koopman Invariant Subspace, i.e. a set of observation functions on which the restriction of the Koopman operator is finite-dimensional, and which gives a good view of the general dynamical system.

\begin{figure}
    \centering
    \includegraphics[width=8.5cm]{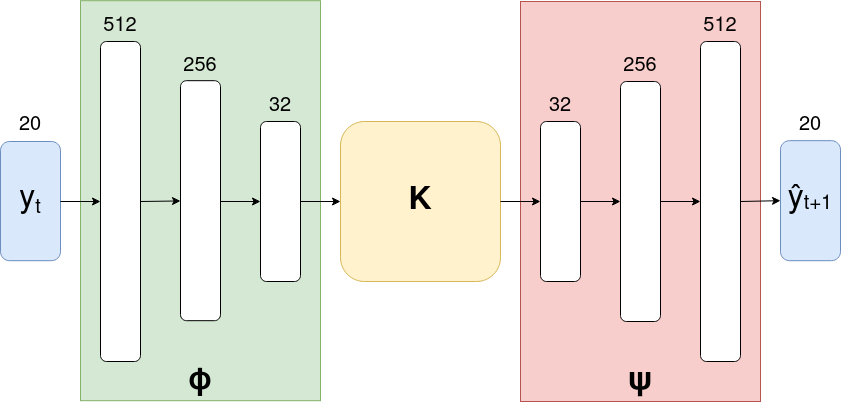}
    \caption{Schematic view of our architecture. Though we precisely represent the number and size of the linear layers from the network on which we experiment, those characteristics could change as long as $\phi$, $\mathbf{K}$ and $\psi$ keep their respective roles. The observation state is of dimension 20 since it contains the reflectances of 10 spectral bands along with their respective derivatives.}
    \label{fig:KoopmanAE}
\end{figure}

We use the neural Koopman architecture from \cite{ICASSP}, which we represent graphically in Figure \ref{fig:KoopmanAE}. In short, this architecture has 2 components: a deep autoencoder ($\phi, \psi$) and a Koopman matrix $\mathbf{K}$. The matrix $\mathbf{K}$, whose entries are trainable parameters, multiplies vectors from the latent space obtained by training the encoder $\phi$ and the decoder $\psi$. It has the effect of advancing time. In terms of equations, this could be written as
\begin{equation}
\label{KoopmanPrediction}
    \psi(\mathbf{K}^{\tau}\phi(\mathbf{x}(t))) = \mathbf{x}(t+ \tau)
\end{equation}
for a given variable $\mathbf{x}$ of time evaluated at a specific time $t$ and advanced by a time $\tau$. Note that a time of 1 classically corresponds to a time step from the time series which is considered (assuming it is regularly sampled).

In the case of satellite image time series, as a first approach, we treat pixels independently from one another. 
Thus, given a time series of $T$ images each containing $N = H \times W$ pixels we denote our state variable as $\mathbf{x}_{i,t}$, where $1 \leq i \leq N$ is the spatial index and $1 \leq t \leq T$ is the temporal index.

Note that, in our case, $\mathbf{x}_{i,t}$ is not a scalar value but a multispectral pixel, i.e. a $L$-dimensional vector, where each of the $L=10$ dimensions corresponds to the reflectance measured for one of the Sentinel-2 spectral bands.
We augment the observation space with the local discrete temporal derivatives of $\mathbf{x}$, which means that we work on data $\mathbf{y}$ defined by
\begin{equation}
\label{y}
    \mathbf{y}_{i,t} = 
    \begin{pmatrix}
    \mathbf{x}_{i,t+1} & \mathbf{x}_{i,t+1} - \mathbf{x}_{i,t}
    \end{pmatrix}
    ^T.
\end{equation}
This is equivalent to the knowledge of the last 2 states of $\mathbf{x}$, and it can therefore be motivated by Takens' embedding theorem \cite{Takens}, which roughly states that the state space gets more predictable when augmented with lagged states. 
Intuitively, it seems much easier to estimate the next step of x when one knows both the current state and its derivative.
$\mathbf{y}$ is now of dimension $2L=$ 20: 10 dimensions for the covered spectral bands and 10 for their derivatives, as shown on Figure \ref{fig:KoopmanAE}.

As in \cite{ICASSP}, we train a prediction model in two stages: first we train it to short-term prediction of the dynamics, i.e. up to 5 time steps 
ahead, and then to long-term prediction, i.e. up to 100 steps 
ahead. 
It is crucial to obtain a model that is able to make good predictions over several years, yet the long-term optimisation problem is highly nonconvex, usually leading to a poor local minimum.
Therefore, the easier short-term prediction task provides a warm-start initialization, avoiding bad local minima. Such a procedure is related to curriculum learning \cite{CurriculumLearning}, which we believe to be crucial when learning difficult physics-related tasks (see \cite{CurriculumLearningSurvey} for a recent survey).

We use 3 different types of loss terms during our training. The main one is the prediction loss $L_{pred}$, which directly represents the $L_2$ distance between the model predictions and the groundtruth. The linearity loss $L_{lin}$ is the $L_2$ distance between the predicted latent vector and the encoding of the actual future state: it ensures that the dynamics is linear in the latent space. The orthogonality loss $L_{orth}$ is a regularization term which encourages $\mathbf{K}$ to be close to an orthogonal matrix, which favors long-term stability as explained in \cite{ICASSP}. Denoting $\Theta$ the set of parameters of our model, i.e. the concatenation of 1) the coefficients of $\mathbf{K}$, 2) the parameters of $\phi$ and 3) the parameters of $\psi$, these loss terms can be written as:
\begin{align}
\label{mid_term}
&L_{pred, \tau}(\Theta) = \sum_{
\substack{
1 \leq i \leq N \\
1 \leq t \leq T-\tau-1
}}  
||\mathbf{y}_{i,t+\tau} - \psi(\mathbf{K}^{\tau}\phi(\mathbf{y}_{i,t}))||^2 \\
&L_{lin, \tau}(\Theta) = \sum_{
\substack{
1 \leq i \leq N \\
1 \leq t \leq T-\tau-1
}}  
||\phi(\mathbf{y}_{i,t+\tau}) - \mathbf{K}^{\tau}\phi(\mathbf{y}_{i,t})||^2 \\
&L_{orth}(\mathbf{K}) = ||\mathbf{KK}^T - \mathbf{I}||_{F}^2
\end{align}

where $||.||_F$ is the Frobenius norm. Note that $L_{pred,0}$ is a classical auto-encoding or reconstruction loss.
Using these basic bricks and setting $\tau_1=5$, $\tau_2=100$, we build our short-term and long-term loss functions as:
\begin{multline}
\label{L_short}\hspace*{-.15cm}
    L_{short}(\Theta) = \beta_1 L_{orth}(\mathbf{K}) + L_{pred,0}(\Theta)  \\
    + L_{pred,1}(\Theta) + L_{pred,\tau_1}(\Theta) + L_{lin,1}(\Theta) + L_{lin,\tau_1}(\Theta)
\end{multline}
\vspace{-20pt}
\begin{multline}
\label{L_long}\hspace*{-.15cm}
    L_{long}(\Theta) = \beta_2 L_{orth}(\mathbf{K}) 
    + \sum_{\tau=0}^{\tau_2}
    (L_{pred,\tau}(\Theta) + L_{lin,\tau}(\Theta))
\end{multline}
One could want to just learn to predict from time 0, which is what is done by the $L_2$ loss in \cite{ICASSP}. This approach
results in a non-robust model which makes good predictions from time 0 but
struggles to make predictions from a different initial time.

So far, we only treated the pixels independently from each other. We now present a simple method that enables to exploit the spatial information of the data.
We use a trained model with frozen parameters
to make long-term predictions from $\mathbf{y}_{.,1}$ using \eqref{KoopmanPrediction}, and assemble the pixel predictions into image predictions $\hat{X}_t \in \mathbb{R}^{H\times W\times L}$ for time t. Using the groundtruth images $X_t$, one can train a convolutional neural network (CNN) to learn the residual function $r: \mathbb{R}^{H\times W\times L} \to \mathbb{R}^{H\times W\times L}$ such that $r(\hat{X}_t) = X_t - \hat{X}_t$. Then, one can add the output of this CNN to a test predicted image to get it closer to the groundtruth. The convolutional layers are expected to partially correct the spatial imperfections made by the pixelwise model.

\section{Presentation of the datasets}
\label{Data}

We selected two areas of interest in France: the forest of Fontainebleau and the forest of Orléans, which are large forestial areas in a region which is moderately cloudy. 
The forest of Fontainebleau in particular has already been studied in remote sensing \cite{Fontainebleau1} \cite{Fontainebleau2}.
Also, since the two sites are separated by about 60 kilometers, one can test a model's transferability by predicting the dynamics of one area after having been trained only on the other one.

The pre-processing steps are largely inspired from the previous work of \cite{joaquim}, although we gathered much more data, both in the spatial and temporal dimensions.
We retrieve the 10m and 20m resolution bands from the Sentinel-2 images with L2A (Bottom Of Atmosphere) correction and perform an imagewise bicubic interpolation on each of the 20m resolution bands to bring all the data to a 10m resolution.

Although the revisit time is only 5 days, 
we identify the images that feature too many clouds and remove elements from the time series accordingly.
This results in
an incomplete time series, where about three quarters of the images have been rejected.
To obtain complete time series, we performed temporal Cressman interpolation \cite{Cressman} with Gaussian weights of radius (i.e. standard deviation) $R$ = 15 days.

\begin{figure}
    \centering
    \includegraphics[width=8.5cm]{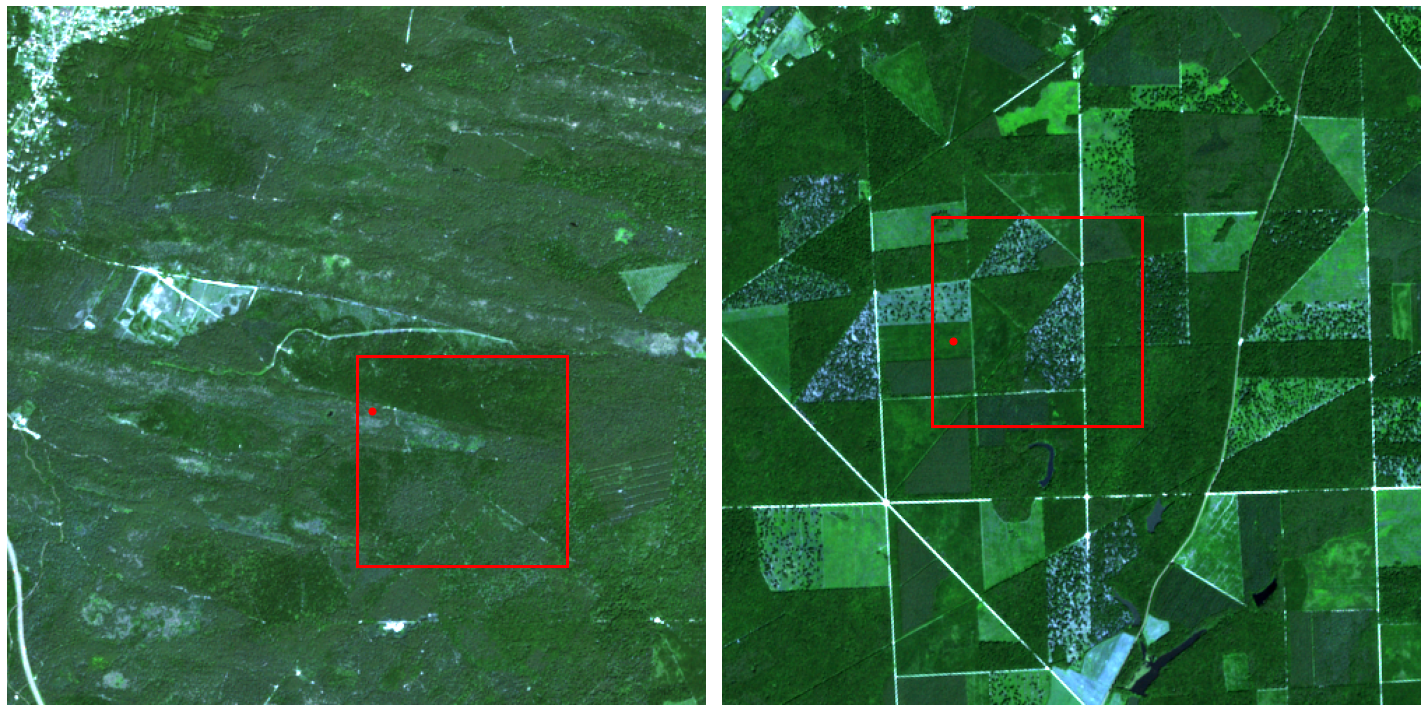}
    \caption{Left: a temporally interpolated Fontainebleau image. Right: a non-interpolated Orléans image. The date for both images is 20/06/2018. Those are RGB compositions with saturated colors. The red squares indicate the $150 \times 150$ pixel subcrops on which we experiment in Section \ref{Experiments}, and the red dots mark the pixels involved in figures \ref{fig:pred_Fontainebleau_graph} and \ref{fig:interpolation_Orléans}.}
    \label{fig:data_samples}
\end{figure}

In the end, we find ourselves with 2 image time series, each of length $T=343$ and image size $500 \times 500$. Given the temporal and spatial resolution of the Sentinel-2 satellites, this corresponds to a a time span of nearly 5 years and to an area of 25 km² each. We also extracted irregular versions of these datasets where no temporal Cressman interpolation has been performed. We show sample images in Figure \ref{fig:data_samples}.

\section{Experiments}
\label{Experiments}


We use a subcrop of $150\times 150$ pixels from the Fontainebleau image time series. The first $T_{train} = 242$ images are used for training and the last $T_{val} = 100$ ones are kept for validation. We extract another $150\times 150$ subcrop from the Orléans time series and use it as a test set.
We train a Koopman autoencoder using successively (\ref{L_short}) and (\ref{L_long}). As shown on figure \ref{fig:KoopmanAE}, the latent dimension of our network is $k=32$.

\subsection{Temporal extrapolation on the training area}
\label{PredFontainebleau}

\begin{figure}
    \centering
    \includegraphics[width=8.5cm]{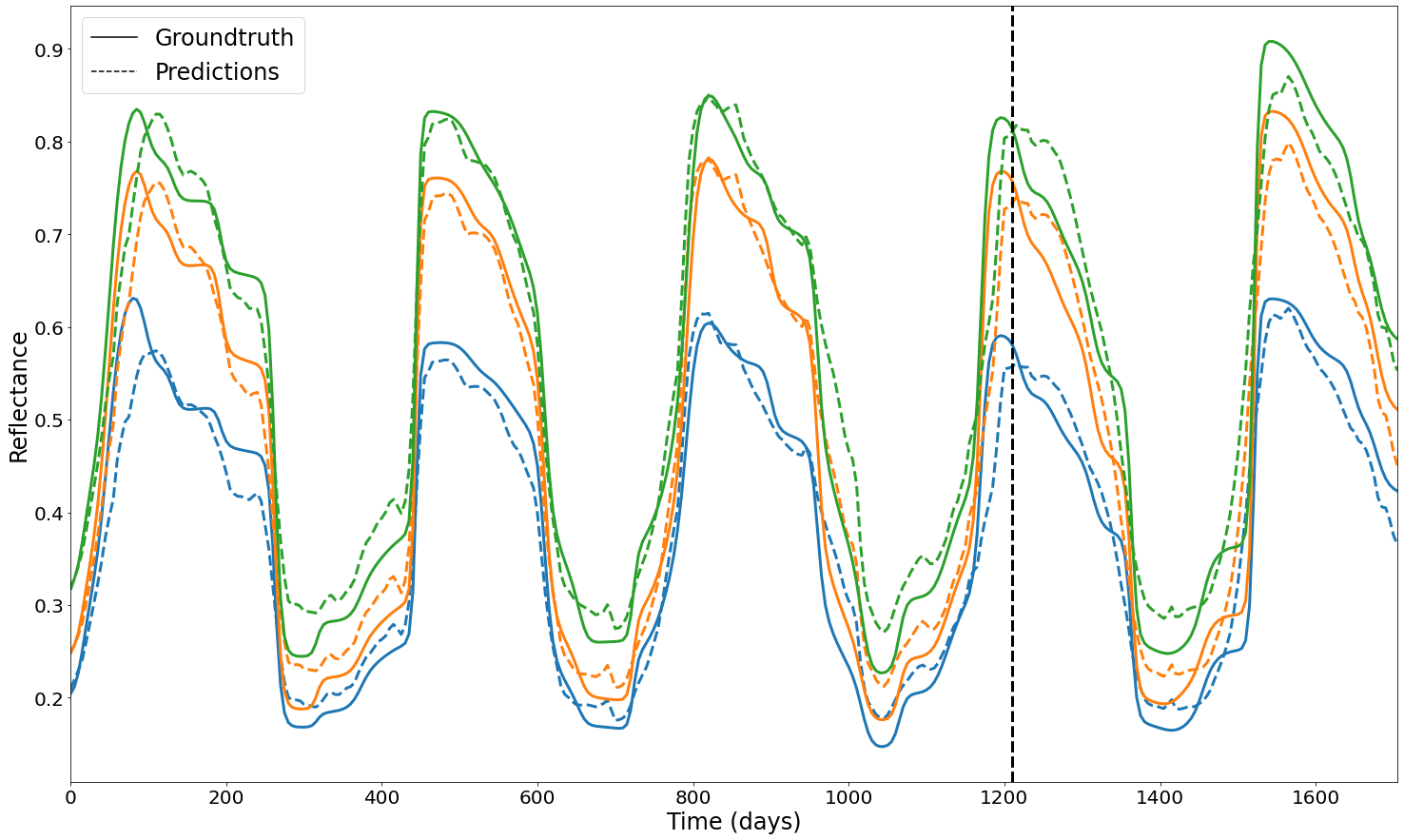}
    \caption{Long-term prediction of reflectances from time 0 for a single pixel from the forest of Fontainebleau, along with the groundtruth. Blue, orange and green respectively denote the B6, B7 and B8A bands. The vertical line marks the separation between the training and validation data.}
    \label{fig:pred_Fontainebleau_graph}
\end{figure}

We first check the ability of our model to extrapolate in time on the Fontainebleau area.
We use the first element of the augmented time series $\mathbf{y}$ from \eqref{y} to
make a $(T_{train}+T_{val})$-time steps prediction, from which the first $T_{train}$ elements correspond to training data while the last $T_{val}$ ones correspond to frames unseen during training. We measure the mean squared error (MSE) between the last $T_{val}$ predicted states and the actual validation data, averaged over all frames, pixels and spectral bands. 
We show an example of such prediction for a random pixel in figure \ref{fig:pred_Fontainebleau_graph}. 

\begin{figure}
    \centering
    \includegraphics[width=8.5cm]{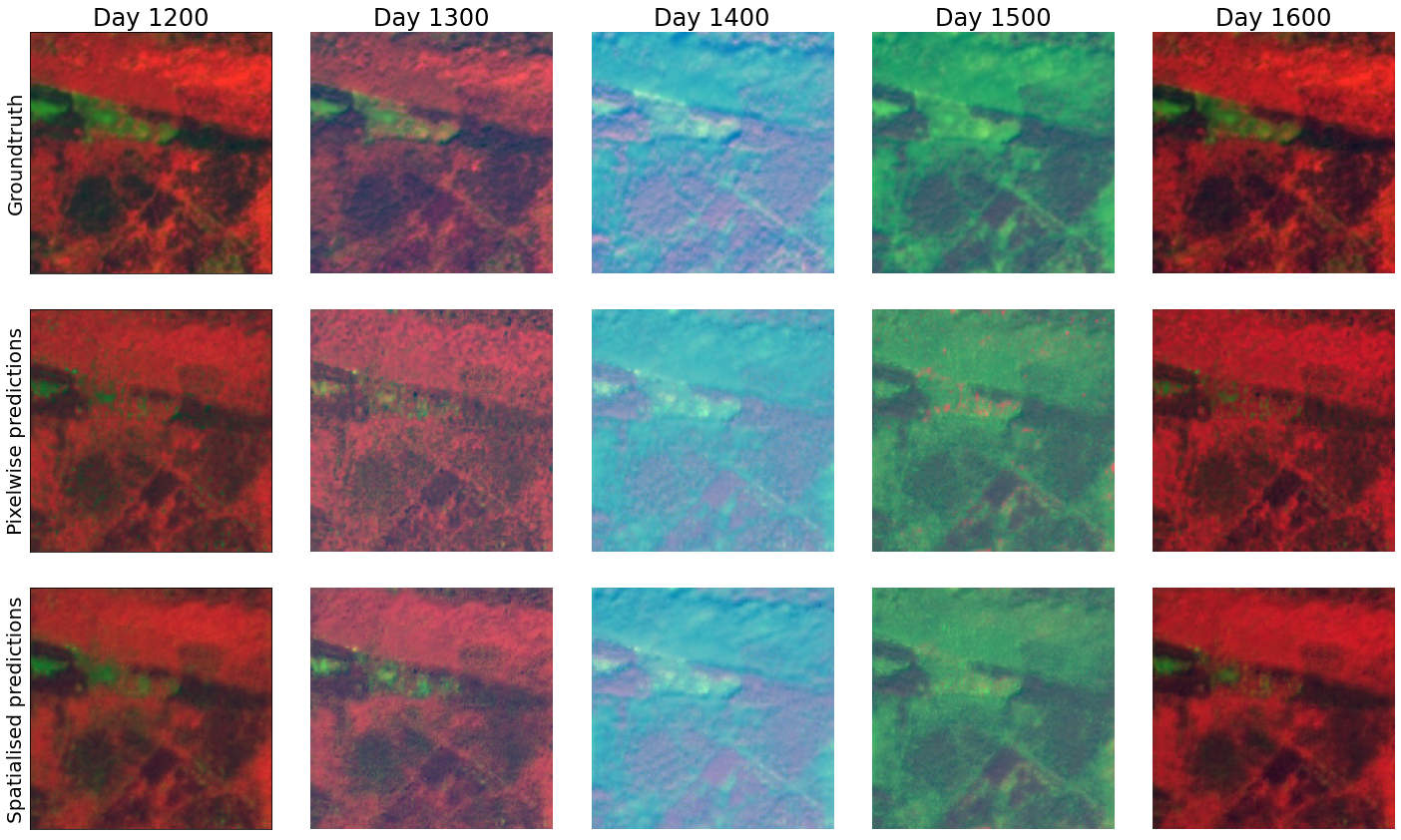}
    \caption{Top: groundtruth images of Fontainebleau, corresponding to test times. Middle: predictions made by our model from state at day 5. Bottom: correction of the middle images by a CNN trained on the pixelwise predictions up to day 1200.
    The colors result from a 3-dimensional principal component analysis (PCA) of the 10 spectral bands performed 
    globally on all the Fontainebleau data.
    This is much more informative than an RGB composition.}
    \label{fig:pred_Fontainebleau}
\end{figure}

We now train a CNN on top of our Koopman model as described in Section \ref{Method}.
We use predictions up to time span $T_{train}$ to train the CNN and then test it on the last $T_{val}$ time steps.
The CNN architecture is very basic, with just 5 convolutional layers and no pooling. The filter sizes are all $3 \times 3$ and the numbers of filters of the successive layers are 64, 64, 32, 32 and 10, totaling 79114 parameters.
As reported in table \ref{prediction_results}, the CNN correction
results in a significant improvement. This can be best visualised when plotting images of the entire predictions, as in Fig. \ref{fig:pred_Fontainebleau}.
One can see that the pixelwise predictions have spatial artifacts in the form of a weaker spatial structure, which is not the case after the CNN correction. Notably, the small area which 
always appears green in the top row of Figure \ref{fig:pred_Fontainebleau},
corresponding to a clearing in the forest, is not well reconstructed by the pixelwise prediction, but this problem is partially addressed by the CNN.
\subsection{Data assimilation on training data}
\label{DAFontainebleau}

The experiment presented in the last subsection shows that our model is indeed able to reconstitute an entire pixel's dynamics from only
an initial condition.
However, this intuitively seems like a difficult task, while using multiple data points to understand a pixel's dynamics seems easier.

We confirm this intuition by a new experiment: using a learned model, 
we look for the latent initial condition $\mathbf{z}_1^*$
from which the propagation by the model best corresponds to the training data.
Formally, for a given spatial index $i$, we seek
\begin{equation}
\label{DAequation}
    \mathbf{z}_1^* = \underset{z_1 \in \mathbb{R}^k}{\textrm{arg min}}\sum_{t=1}^{T_{train}} ||\mathbf{y}_{i,t} - \psi(\mathbf{K}^{t-1} \mathbf{z}_1)||^2. 
\end{equation}

We emphasize that, here, only the latent initial condition varies while the model parameters remain fixed. This is a kind of variational data assimilation \cite{VariationalDA} where everything is based on the data, since the model itself has been trained fully from the data. Finding the best initial condition is done by a gradient descent which backpropagates into the whole pretrained model. This optimisation problem is not convex, yet starting from a null initial latent state gives satisfactory results, and starting from the encoding of the actual initial state gives even better ones. 

When making predictions using the result of the gradient descent as the initial latent state, not only do we fit the assimilated data very well, but we also obtain excellent extrapolations. As can be seen in Table \ref{prediction_results}, the MSE is far lower than when predicting from only one data point.

\subsection{Data assimilation on test data}
\label{DAOrléans}

We now move on to the Orléans site, 
from which no data has been seen during training,
and we aim at transfering the knowledge of the Fontainebleau area without training a new model.
The change of area results in a data shift, to which the task of prediction from a single reflectance vector (like in subsection \ref{PredFontainebleau}) is very sensitive, leading to relatively poor results with our model trained on Fontainebleau. However, when 
performing variational data assimilation
as in section \ref{DAFontainebleau}, one can perform a good prediction without even needing a complete time series to do so. Indeed, our model can easily handle irregular data, and in our tests it has even been more effective to do so than to
assimilate on an interpolated time series.
The only difference is that one should only compute the prediction error on the time indexes from the set $S \subset \{1,2,...,342\}$ of available data, i.e. rewrite (\ref{DAequation}) as
\begin{equation}
    \mathbf{z}_1^* = \underset{\mathbf{z}_1 \in \mathbb{R}^k}{\textrm{arg min}} \sum_{t \in S} ||\mathbf{y}_{i,t} - \psi(\mathbf{K}^{t-1} \mathbf{z}_1)||^2. 
\end{equation}

We consider a set of 94 irregularly sampled images from the forest of Orléans, each with its associated timestamp, over the same time interval as the training and validationov data.  
We intentionally kept some partially cloudy data in this set. 

\begin{figure}
    \centering
    \includegraphics[width=8.5cm]{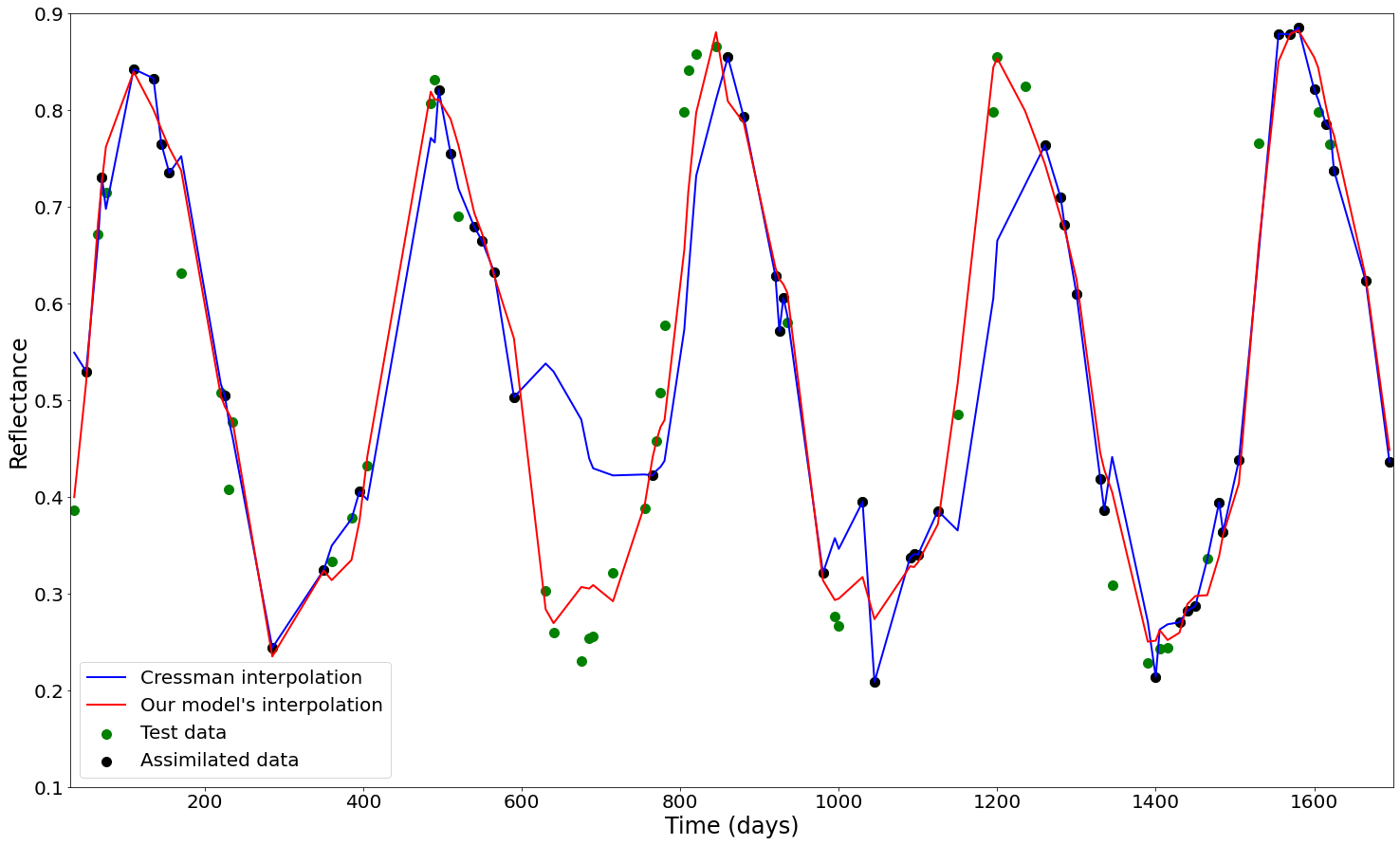}
    \caption{Comparison of interpolations for an Orléans pixel on the B7 band, using a Cressman method and using data assimilation with our model trained on Fontainebleau data.}
    \label{fig:interpolation_Orléans}
\end{figure}

First, we test our model in a classical data assimilation setting, where we check that it is able to interpolate from some of the data to recover the part of the data that was kept aside. We check that our method does better than a well-parameterized Cressman interpolation.
The setup is the following: for each image, we keep it with a probability $0.5$. We then interpolate on the retained images and use the MSE on the removed images as the performance measure. 
We perform a Gaussian Cressman interpolation with radius 
$0.5,1,...,6.5,7$ 
time steps (i.e. 2.5 to 35 days) and compare the best result to the data assimilation method with our model. 
We repeat this experiment with 6 different sets of retained images, looking for the best performing Cressman parameter at each iteration, and average the results.
Our method always outperformed the best Cressman interpolation by a margin of at least 25\%. The average MSE obtained by the Cressman interpolation was $5.72 \times 10^{-3}$, and the one from our model was $3.36 \times 10^{-3}$. One can visually assess the quality of our interpolation on figure \ref{fig:interpolation_Orléans}, and see that the model
was able to combine
the information from different years to recover the correct periodic pattern, ignoring the noisiest data points.

\begin{figure}
    \centering
    \includegraphics[width=8.5cm]{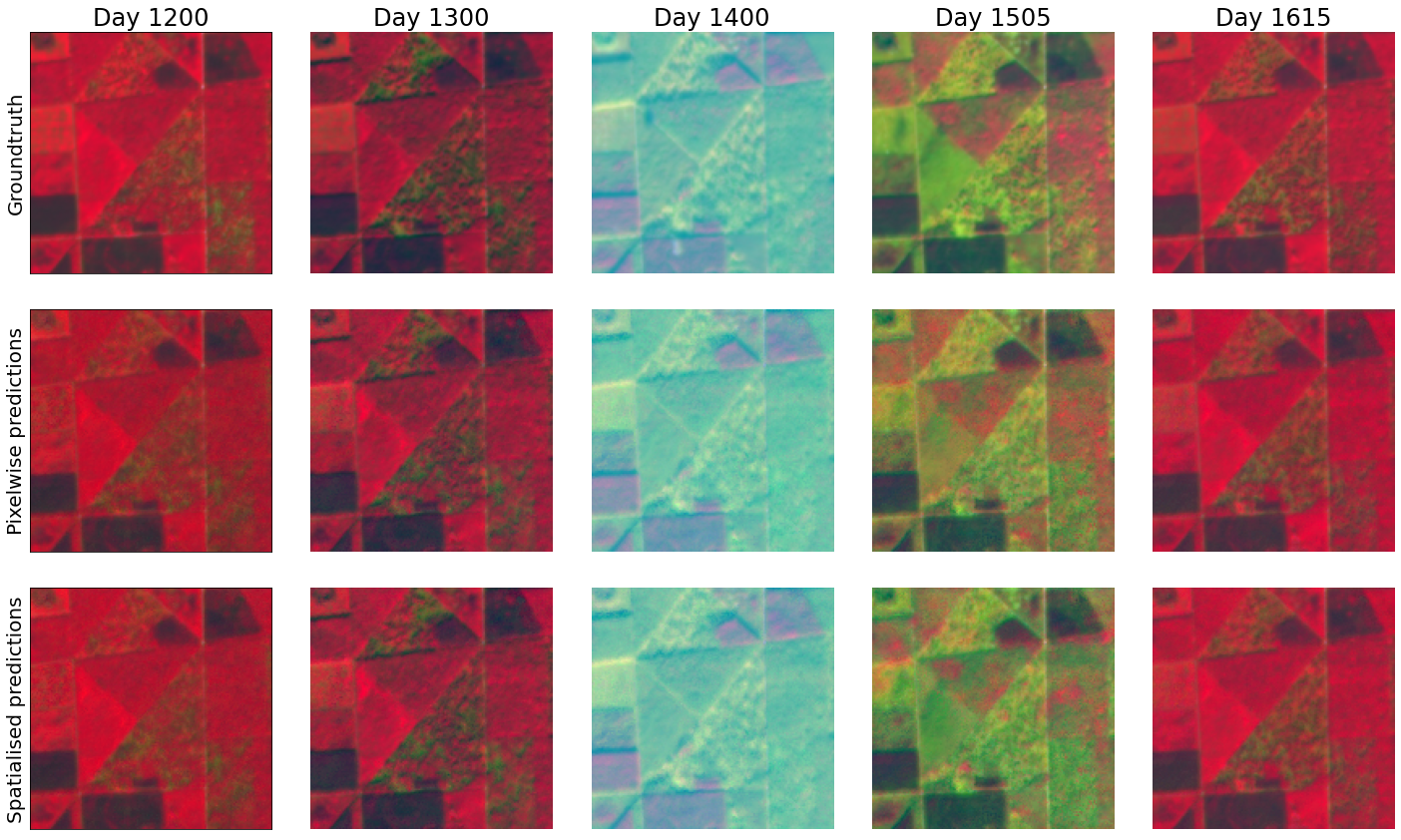}
    \caption{Top: groundtruth images of Orléans, corresponding to test times. Middle: predictions made by our model, assimilated on irregularly-sampled earlier images. Bottom: correction of the middle images by a CNN trained on the assimilated data. Like in Figure \ref{fig:pred_Fontainebleau}, the colors are obtained from a 3-dimensional PCA of the Orléans data.}
    \label{fig:DA_extrapolation_Orléans}
\end{figure}

We now perform forecasting using the same method as in Section \ref{DAFontainebleau}. We keep the last 31 images to test the prediction performance, and perform data assimilation on the remaining images. Some results can be observed on Figure \ref{fig:DA_extrapolation_Orléans}.

\subsection{Discussion of the results}

Our prediction performances are synthesized in Table \ref{prediction_results}. Note that the Fontainebleau data is an interpolated regular time series while the Orléans data corresponds to irregularly-spaced data points with no temporal interpolation. 

One can observe that performing data assimilation with several data points is generally far more effective than performing a prediction from a single data point at time 0.
Although all of our methods perform far worse on the data from the forest of Orléans than on the training area in the forest of Fontainebleau, the usage of data assimilation partially mitigates the shift in the data. One can conjecture that, although the pseudo-periodic pattern of the reflectance dynamics does not depend on the initial condition in the same way in the Orléans data than in the Fontainebleau data, the model can still identify a known pattern when fed with more data from an Orléans time series.

Overall, backpropagating through a long time series prediction is easy because of the simplicity of our model: predicting one step ahead only costs one matrix-vector multiplication, and the most computationally intensive part of the prediction is actually the encoding and decoding of data.

\begin{table}[htbp]
\caption{Forecasting performance of our prediction models}
\begin{center}
\begin{tabular}{|c|c|c|}
\hline
\textbf{Method} & \textbf{\textit{Fontainebleau}}& \textbf{\textit{Orléans prediction MSE}} \\
& 
\textbf{\textit{prediction MSE}}& \\
\hline
Prediction from time 0 & $1.87 \times 10^{-3}$ & $7.13 \times 10^{-3}$  \\
\hline
Prediction from time 0 & $1.37 \times 10^{-3}$  & $4.23 \times 10^{-3}$ \\
with CNN correction &  & \\
\hline
\hline
Prediction with & $2.89 \times 10^{-4}$ & $1.15 \times 10^{-3}$ \\
data assimilation & & \\
\hline
Prediction with & $\mathbf{2.79 \times 10^{-4}}$ & $\mathbf{1.07 \times 10^{-3}}$ \\
data assimilation & & \\
and CNN correction & & \\
\hline
\end{tabular}
\label{prediction_results}

\end{center}
\end{table}
\section{Conclusion}
\label{Conclusion}

We showed an adaptation of the previously introduced method from \cite{ICASSP} to real satellite image time series, in order to learn an unsupervised model which is able to perform several downstream tasks even using irregular data. 
Note that our assimilation experiment was a very simple proof of concept since only the initial latent state was optimized using a frozen model, yet one could also imagine a variational data assimilation procedure in which the model parameters are allowed to vary.
More generally, there are many downstream tasks in which our model might be of use, e.g. classification tasks in few-shot settings.
A natural extension to this work would be to show the model ability to learn from more difficult data, for example with a higher diversity of images, e.g. different crop types and urban environments, with diverse underlying dynamic patterns. One could also test the ability of our model to handle complex spatio-temporal missing data patterns. 
In particular, although we demonstrated the ability of our trained model to handle irregular test data, the training was still performed on regular data.
A weakness of our method is that
most of the computation is done pixelwise, 
and the spatial structure of the data is only used a posteriori through a CNN model. It might be of interest to encode some spatial information directly in the Koopman autoencoder. 
Other possible extensions include the ability to exploit a control variable or to provide uncertainties along with the predictions.


\end{document}